\documentclass[preprint,showpacs,superscriptaddress,floatfix,prl]{revtex4}

\usepackage{amsmath}
\usepackage[dvips]{graphicx}
\usepackage{epsfig}
\usepackage{tabularx}
\usepackage{color}
\usepackage[pdfpagemode=UseNone,colorlinks=true,linkcolor=blue,citecolor=blue]{hyperref}

\begin{document}

\title{Levitated optomechanics with a fiber Fabry-Perot interferometer}
%AAG added author
\author{A. Pontin$^{\star}$}
\author{L.S. Mourounas}
\affiliation{\small{
Physics Department, University College London, London, UK\\
$\star$ e-mail:  {a.pontin@ucl.ac.uk}
}}
\author{A.A. Geraci}
\affiliation{\small{
Physics Department, University of Nevada, Reno, NV, USA\\
}}
\author{P.F. Barker}
\affiliation{\small{
Physics Department, University College London, London, UK\\
$\star$ e-mail:  {a.pontin@ucl.ac.uk}
}}

%\date{\today}
\begin{abstract}
In recent years quantum phenomena have been  experimentally demonstrated on variety of optomechanical systems ranging from micro-oscillators to photonic crystals. Since single photon couplings are quite small, most experimental approaches rely on the realization of high finesse Fabry-Perot cavities in order to enhance the effective coupling. Here we show that by exploiting a, long path, low finesse fiber Fabry-Perot interferometer ground state cooling can be achieved. We model a $100$~m long cavity with a finesse of $10$ and analyze the impact of additional noise sources arising from the fiber.   As a mechanical oscillator we consider a levitated microdisk but the same approach could be applied to other optomechanical systems.
\end{abstract}

%\pacs

\maketitle
Cavity optomechanics\cite{akm} has achieved several long-awaited experimental results highlighting the quantum nature of the interaction. From the generation of ponderomotive squeezing\cite{brooks,painter,purdy} and field quadrature QND measurement\cite{my1} to the cooling of the mechanical motion to a thermal occupation number below unity\cite{connel,teufel,chan,harris}. These results, obtained in a variety of systems, have increased the interest in the generation of other non-classical states and in the investigation of the quantum to classical transition. In recent years, optical cooling of levitated dielectric nanoparticles\cite{nanop} has been receiving a lot of attention. These unclamped oscillators offer the possibility to be operated in a regime where thermal noise, due to the residual background gas, is not the main contribution to the overall decoherence rate. Typically, the nanoparticle is trapped by optical tweezers\cite{twee} or an electro-dynamic\cite{Ptrap} trap and cooled by an optical cavity field. In these configurations random momentum kicks to the nanoparticle associated with radiation pressure shot noise represent a major limitation toward ground state cooling, as has been recently reported\cite{novotny}.

An intriguing possibility towards the suppression of recoil heating is to levitate an apodized microdisk. If its radius is significantly bigger than the optical waist a microdisk behaves as a thin dielectric slab for which scattering occurs only due to surface roughness. This is in stark contrast to a sub-wavelength nanosphere that scatters light in a dipole field pattern. A similar system was initially prosed in Ref.~\cite{chang2}, where a tethered microdisk was considered. They showed that by apodizing the edges of the microdisk even for a radius comparable to the waist, the scattering limited finesse is $\gg10^4$.

\begin{figure}[h]
\begin{center}
  \includegraphics[width=0.47\textwidth]{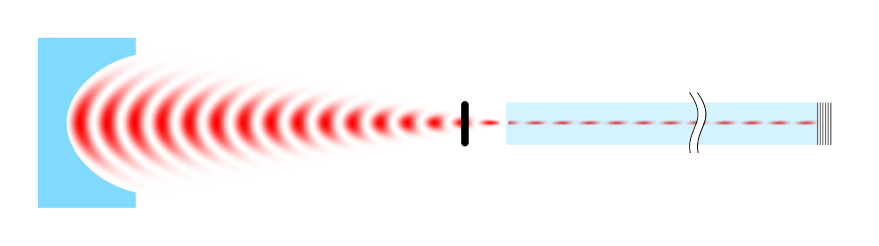}
  \caption{\label{scheme1} Scheme of the FFP interferometer. The optical cavity is divided in two parts. A free space region where the microdisk is trapped and an optical fiber. The optical mode transition from a guided HE11 mode to a Gaussian mode. }
\end{center}
\end{figure}

Most optomechanical systems require a high finesse optical cavity in order to enhance the light matter interaction. Here, we propose a levitated microdisk trapped in the standing wave of a long low finesse extrinsic fiber Fabry-Perot (FFP) interferometer. This scheme is shown in Fig.~\ref{scheme1}. The input field is injected into the cavity via an input coupler with a small radius of curvature, the field is propagated in free space for a few millimeters and then coupled into a single mode fiber.  At the far end of the fiber a high reflectivity mirror or a distributed Bragg reflector provides the end mirror for the FFP.
%\begin{figure}[h]
% \centering
%\includegraphics[width=0.47\textwidth]{hgfadecc.png}
%\caption{(Color online) Sketch of the different stages of the experimental configuration, from the oscillator in a thermal state (a) to the bright squeezed state (e). The explanation is in the main text.}
% \label{funzionamento}
%\end{figure}

There are three critical aspects that need to be addresseed. These include the optical losses that are introduced at the fiber/free space interface, the cavity mode volume that will determine the microdisk coupling to the optical fields and the additional noise sources and non-linear effects introduced by the fiber that could hinder the overall performance of the system.

\begin{figure}[h]
\begin{center}
  \includegraphics[width=0.47\textwidth]{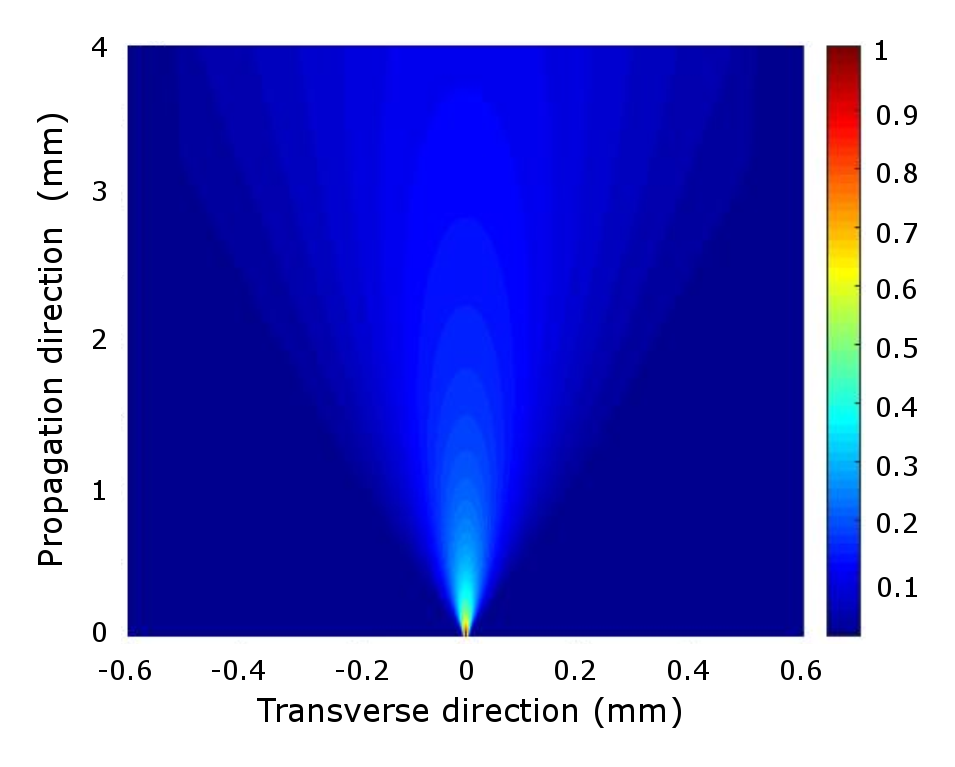}
  \caption{\label{propagation} Intensity distribution of HE11  mode propagated in free space.  }
\end{center}
\end{figure}

\textbf{Optical losses} have been evaluated with numerical methods aimed at calculating the cavity reflection coefficient (considering ideal input and output couplers). The beam was propagated from the fiber tip in free space using a finite difference beam propagation method exploiting the assumption of slowly varying fields\cite{young}. Therefore, the initial field is a HE11 mode. After a length $L_{free}$ the beam was reflected by a mirror  and propagated back to the fiber. The beam was than propagated through $1$~mm of fiber implementing a  propagation method\cite{feit}. The total round trip power loss is obtained by comparing the initial and final power. The parameters considered are $L_{free}=4$~mm, a wavelength of $\lambda =1550$~nm and a Corning SMF-$28$ optical fiber; with these values an overall power loss of $4.13$~\% was calculated, corresponding to an interface limited cavity finesse of $\mathcal{F}\simeq150$. An example of the intensity profile obtained before reflection in shown in Fig.~\ref{propagation}.

\textbf{The cavity mode volume} is defined as
\begin{equation}\label{modevolume}
  V_m=\int|E(\textbf{r})|^2 dV
\end{equation}
\noindent where $E(\textbf{r})$ is the normalized cavity field.
%AAG should we specify this is a normalized (dimensionless) cavity field?
We divide the integral in two domains, fiber and free space. In the former $E(\textbf{r})=cos(k z) Exp(-r^2/w_o^2)$, where $k=2\pi/\lambda$ and $w_o$ is the fiber mode field radius (mfr), while in the trapping region
\begin{equation}\label{modeprofile}
E(\textbf{r})=cos(k z) \frac{Exp(-r^2/w^2(z))}{\sqrt{1+(z+z_R)^2}}
\end{equation}
\noindent with $w(z)=w_o\sqrt{1+(z/z_R)^2}$ where $z_R$ is the Rayleigh range. Since the beam is highly divergent Eq.~\ref{modeprofile} is an estimation. To be correct one needs to include the curvature of the wavefronts and consider the details of the mirror geometry. However, for the parameters considered in the following the para-axial approximation holds\cite{Tey} and the contribution to the total mode volume coming from the free space region is only of the order of a few \% thus Eq.\ref{modeprofile} provides a good estimate. By evaluating the integral in Eq.~\ref{modevolume} we find
%\begin{equation}\label{modevolumefinal}
%V_m=\frac{\pi w_o^2 L_{free}}{8}\left(1+\frac{L_{free}^2}{12 z_R^2}\right)+\frac{\pi w_o^2 n_f L}{4}
%\end{equation}
\begin{equation}\label{modevolumefinal}
V_m=\frac{\pi w_o^2 n_s L}{4}\left(1+ \frac{L_{free}}{ n_f L}  \right)
\end{equation}
\noindent where $L_{free}$ is the length of the free space region, L is the fiber length and $n_s$ its refractive index. A fiber cavity allows to achieve a cavity waist of the order of the wavelength without the need work in the near concentric configuration close to the instability region\cite{durak}.

%For a fiber cavity the reduction of the mode volume compared to a standard Fabry-Perot cavity of same length and waist $w_c$ is given by $n_f (w_o/w_c)^2$ %(Assuming that the mode volume is $\pi w^2 L/4$ even in the concentric configuration).

\textbf{Fiber noises and non-linear effects}. We are going to assume that the environmental, electronic and classical laser noises can be controlled to a negligible level, then, the fundamental noise introduced by the fiber is the thermoptic induced phase noise, usually referred to as thermal phase noise in the fiber community. Since the intensities required for trapping the microdisk are typically rather high, possible issues could also arise from non-linear effects like Brillouin and Raman scattering.

\textit{Fiber Thermal noise}. Fiber interferometers, in various configurations(Mach-Zehnder, Michelson...), constitute an active field of research especially for sensing applications\cite{fibersensors}. The current generation of devices are approaching the fundamental thermal noise limit. This has been measured with high accuracy in a Mach-Zehnder interferometer\cite{bartolo} and compared to a model initially proposed by Wanser\cite{wan}. In his theory the power spectral density (PSD) of phase noise for a fiber of length $L$ can be estimated to be\cite{dong}
\begin{equation}\label{psdphase}
S_{\phi\phi}(\omega)=\pi\frac{L k_B  T^2}{\kappa_t}\left(\frac{n_s q}{\lambda}\right)^2 F(\omega)
\end{equation}
\noindent where $q=\alpha_L+\frac{1}{n_s}\frac{dn_s}{dT}$ is the thermoptic coefficient, $\alpha_L$ the  linear expansion coefficient, $\kappa_t$ is the thermal conductivity of the fiber medium and $F(\omega)$ is a term that characterizes a frequency cut-off dependent on  fiber geometry. It is given by:
\begin{equation}\label{fomega}
 F(\omega)=ln\left(\frac{k_{max}^4+(\omega/D)^2}{k_{min}^4+(\omega/D)^2}\right).
\end{equation}
In this expression $k_{max}=2/w_o$, $k_{min}=2.405/a_f$, where $a_f$ is the fiber outer radius, and $D$ is the thermal diffusivity.  Eq.~\ref{psdphase} describes the variance of the phase after the light field as passed through the fiber a single time, naturally, in a FFP the light bounces multiple times between the cavity mirrors so that the final total phase noise grows with an increasing finesse. In order to include thermal phase noise in the cavity dynamical equations it is simpler to consider it as detuning noise, that is $S_{\dot{\phi}\dot{\phi}}(\omega)=(c/2 n_s L)^2 S_{\phi\phi}(\omega)$, where $c$ is the speed of light.

\textit{Raman and Brillouin Scattering}. For an optical field propagating in a molecular medium a fraction of the total power can be transferred to a frequency downshifted field through the interaction with the vibrational modes of the medium. Acoustical phonons are involved in Brillouin scattering while optical phonons participate in Raman scattering. For both processes the non linear dynamics becomes exponentially more relevant after a critical threshold is surpassed. In the case of Raman scattering the critical power can be estimated as\cite{agrawal} $P_{cr}\approx  \frac{16 A_{eff}}{ g_R~L}$ where $L$ is the fiber length, $A_{eff}= \pi w_o^2$ is the effective mode area and $g_R\simeq6.4~10^{-14}$~m/W is the peak Raman gain. A typical value for the mode-field radius at $1550$~nm is $5.25~\mu$m and considering a $100$~m long fiber, than $P_{cr}=1$~W. A similar expression can be exploited for the case of Brillouin scattering\cite{agrawal} $P_{cr}\approx  \frac{21 A_{eff}}{ g_B~L}$, where $g_B\simeq 5~10^{-11}$~m/W is the peak brillouin gain. For the parameters considered before we obtain $P_{cr}\simeq350$~mW. As for the case of phase noise, these values correspond to a single pass through the fiber. For a FFP the thresholds can be significantly reduced\cite{ogusu,labudde}. However, lower values for $g_B$ have been reported in the literature\cite{lanticq}. Furthermore, stimulated Brillouin scattering is one of the most important limiting factors in high power fiber lasers and, as such, increasing its threshold is a highly researched topic. The mainstream approach relies on the introduction of non-uniformities in the fiber to achieve spectral broadening of the Brillouin gain spectrum, thus reducing the peak value $g_B$. These  non-uniformities ranges from temperature gradients\cite{kova} to modifications of fiber composition or geometry\cite{koby,evert}.
%\begin{figure*}[t]
%\centering
%\includegraphics[width=1\textwidth]{3D_composed_v3.png}
%\caption{(Color online) Phase space distributions for the three configurations named respectively a), d) and e) in the scheme of Fig.\ref{funzionamento}: from left to right, thermal oscillator (a) at the effective temperature $T_{\mathrm{eff}}\simeq$15 K ($\gamma_{\mathrm{eff}} =$110 Hz); parametrically squeezed oscillator (d), with a parametric gain $g =0.83$; squeezed oscillator with coherent excitation and frequency control (e), with a parametric gain $g =5.4$.}
% \label{3Ddist}
%\end{figure*}

\textbf{Description of the model}. We consider an apodized microdisk, of radius $a$ and thickness $t$,  trapped in the standing wave of the FFP close to the fiber/free-space interface; we assume a high aspect ratio $a/t>w_o/\lambda$ in order to minimize both modifications of the Gaussian profile and scattering of the intra-cavity field. We focus on the center of mass CM degree of freedom of the microdisk along the cavity axis. The transverse confinement is typically weaker giving much lower dynamical timescales while the lowest flexural mode typically has a frequency $>>1$~MHz.
Three beams drive the cavity: a high power trapping beam at $\lambda_{trap}$ and two low power beams at $\lambda_{cm}\simeq\lambda_{trap}\equiv\lambda$ to cool and detect the microdisk motion. The model we are considering is, thus, an extension of that presented in Ref.\cite{tania}. We add to that description an additional field and include the fiber phase noise contribution. It must be pointed out that this treatment is based on the high finesse approximation, that is, describing the optical resonance as a Lorentzian. For the finesse values that we are going to consider the difference with the Airy peak  and a Lorentzian can be significant. The equation of motions are:
\begin{equation}\label{dyn1}
\begin{split}
  \dot{a}_i =& -[\kappa-i ( \Delta_o^i+\dot{\phi}_i)]~a_i +i g_o a_i ~cos^2(k x-\phi_i) \\ & + \sqrt{2\kappa_{in}}~\alpha_{in,i} + v_i \\
  \ddot{x} =& \frac{\xi}{m}-\gamma_g \dot{x}-\frac{\hbar k g_o}{m} \sum_i  a_i^\dag a_i ~sin[2(k x-\phi_i)]
  \end{split}
\end{equation}
\noindent where $i=t,c,m$ meaning trap, cooling and meter fields. In Eqs.~\ref{dyn1} $g_o=\frac{V_d}{2 V_m}(\epsilon-1)\omega_l$ is the coupling strength, $\omega_l$ is the field frequency, $\Delta_o^i$ is the empty cavity detuning, $\kappa=\kappa_{in}+\kappa_{out}+\kappa_{loss}$ is the total cavity half-linewidth, $\alpha_{in,i}$ is the driving amplitude, $v_i=\sqrt{2 \kappa_{in}}~a_{in,i}+\sqrt{2 \kappa_{out}}~a_{out,i}+\sqrt{2 \kappa_{loss}}~a_{loss,i}$ is a weighted sum of all vacuum operators and $\dot{\phi}_i$ is a detuning noise term that accounts for the fiber phase noise. This is considered to provide an uncorrelated contribution to all cavity fields, that is $\langle\dot{\phi}_i(t)\dot{\phi}_j(t')\rangle=0$. The field fluctuations are uncorrelated and have the following correlation functions\cite{zoller}
\begin{equation}\label{corr}
\begin{split}
 \langle a_i(t) a_j(t')\rangle =& \langle a_i^\dag(t) a_j^\dag(t') \rangle =\langle a_i^\dag(t) a_j(t')\rangle =0 \\
  \langle a_i(t) a_j^\dag(t')\rangle =& \delta (t-t') \\
\end{split}
\end{equation}
Finally, $\xi$ is a Brownian stochastic force with zero mean value that arises from the background gas and obeying the correlation function\cite{zoller,landau}:
\begin{equation}\label{thermal}
\langle\xi(t)\xi(t')\rangle=\frac{\gamma_g}{\omega_t}\int\frac{d\omega}{2 \pi} e^{-i\omega(t-t')}\omega \left[ coth \left( \frac{\hbar \omega}{2k_B T}  \right)+1  \right]
\end{equation}
\noindent where $k_B$ is the Boltzman constant and $\gamma_g$ is the viscous damping rate.

We consider $\alpha_{in,c}=R_1~\alpha_{in,t}$ and $\alpha_{in,m}=R_2~\alpha_{in,t}$ with $0<R_1,R_2\leq 1$. The steady state is readily obtained to be
\begin{equation}\label{steady}
\begin{split}
%\nonumber %to remove numbering (before each equation)
 \alpha_{i} &= \frac{\sqrt{2 \kappa_{in}}}{\kappa-i~\Delta^i}~\alpha_{in,i},\\
  -\frac{sin[2(k x_o-\phi_1)]}{sin[2(k x_o-\phi_2)]} &=\frac{(1+\delta_t^2)}{(1+\delta_c^2)(1+\delta_m^2)}\\
  \,\,\,\,&\left[R_1^2(1+\delta_m^2)+R_2^2(1+\delta_c^2)\right]
  \end{split}
\end{equation}
\noindent where $\Delta_i$ is the hot cavity detuning and $\delta_i=\Delta_i/\kappa$. Upon displacement of the operators in Eqs.~\ref{dyn1} and subsequent linearization the dynamical equations become
\begin{equation}\label{dyn2}
\begin{split}
%\nonumber %to remove numbering (before each equation)
  \dot{a}_i =& -(\kappa-i  \Delta_i)~a_i -i g_o k  \alpha_{i} ~sin[2(k x_o-\phi_i)] ~x -i\alpha_{i}\dot{\phi}+ v_i \\
  \ddot{x} =& -\omega_t^2 x -\frac{\hbar k g_o}{m} \sum_i  (a_i^\dag~\alpha_{i}+ a_i~\alpha_{i}^* ) ~sin[2(k x_o-\phi_i)]\\ &-\gamma_g \dot{x}+\xi/m
\end{split}
\end{equation}
\noindent where $\omega_t^2=\frac{2 \hbar k^2 g_o}{m}\sum_{i}\left( |\alpha_{i}|^2 ~cos[2(k x_o+\phi_i)]\right)$ is the optical trap frequency. In the following we will assume $\phi_1=0$, $\phi_2=\pi/4$, $R_1,R_2\ll 1$ and $\Delta_t=\Delta_m=0$ so that $x_o\simeq0$ represents a good approximation considerably simplifying the model since the effective optomechanical parameters are purely determined by the cooling field. Thus, by moving into Fourier space and defining
\begin{equation}\label{rates}
 A_{i,\pm}(\omega)=\frac{\hbar k^2 g_o^2 |\alpha_{i}|^2}{m\omega}\frac{\kappa}{\kappa^2+(\omega \mp\Delta_i)^2}
\end{equation}
\noindent with which the effective mechanical parameters can be expressed as $\gamma_{eff}=\gamma_m+\gamma_{opt}=\gamma_m+A_{c,-} - A_{c,+}$ \footnote{To simplify the notation we use $A_{i,\pm}\equiv A_{i,\pm}(\omega_t)$} and $\omega_{eff}^2=\omega_t^2+\frac{\omega}{\kappa}\left[(\Delta_c+\omega)~A_{c,-} +(\Delta_c-\omega)~A_{c,+}\right]$, the mechanical susceptibility is $\chi_{eff}(\omega)=[m~(\omega_{eff}^2-\omega^2-i\omega\gamma_{eff})]^{-1}$.% Whit these definitions it is easy to verify that
%\begin{equation}\label{disp}
%\frac{x}{\chi_{eff}(\omega)}=-\hbar k g_o \left\{ \sum_i \left[K_i(\omega) \alpha_i^* v_i +K_i(-\omega)^* \alpha_i v_i^\dag +i |\alpha_i|^2 \left(K_i(-\omega)^*-K_i(\omega)\right)~\dot{\phi}_i\right]\right\}+\xi
%\end{equation}
%\noindent where $K_i(\omega)=[\kappa-i(\omega+\Delta_i)]^{-1}$. From Eq.\ref{disp}
 The symmetrized displacement PSD, then, is given by
%AAG should the last term in the equation be summed over "i"?
\begin{equation}\label{disp_psd}
\begin{split}
\frac{\bar{S}_{xx}(\omega)}{|\chi_{eff}(\omega)|^2}=&S_{th}+ \sum_i \left[\hbar m \omega ~(A_{i,+} + A_{i,-})\right] \\&+ \sum_i 4 \Delta_i^2 \frac{m^2 \omega^2}{g_o^2 k^2 \kappa^2}~A_{i,+}~A_{i,-}~S_{\dot{\phi}\dot{\phi}}(\omega).
\end{split}
\end{equation}
\noindent Eq.~\ref{disp_psd} accounts for all force noises acting on the microdisk except for recoil heating due to the trapping potential. This can be included through the substitution $S_{th}\rightarrow S_{th}~(1+\gamma_{sc}/ \bar{n} \gamma_g)$, where $\bar{n}=k_B T/\hbar \omega_t$ is the initial phonon number and $\gamma_{sc}=\frac{V_m}{V_d}\frac{\lambda}{4 L}\frac{\omega_t}{(\epsilon-1)\mathcal{F}_{disk}}$ is the recoil heating rate in which $\mathcal{F}_{disk}\simeq10^5$ is the disk-limited cavity finesse\cite{chang2}. By assuming $\omega_t\gg \bar{n}\gamma_g,g_i$ and $\kappa\gg \gamma_g,g_i$, where and $g_i=g_o k \sqrt{\hbar /m \omega_t} |\alpha_i|$ is the effective coupling strength, the final phonon occupation number is given by\cite{vitali2}
\begin{equation}\label{phonon}
  n_f=\frac{\bar{n}\gamma_g+\gamma_{sc}+\sum_i[A_{i,+}+2\Delta_i^2\frac{m \omega_t}{\hbar g_o^2 k^2 \kappa^2}~A_{i,+}A_{i,-}S_{\dot{\phi}\dot{\phi}}(\omega_t)]}{\gamma_{eff}}
\end{equation}
%\begin{equation}\label{phonon}
%  n_f=\frac{\bar{n}\gamma_g+\gamma_{sc}+A_{m,+}+A_{c,+}[1+2\Delta_c^2\frac{m \omega_t}{\hbar g_o^2 k^2}~A_{c,-}S_{\dot{\phi}\dot{\phi}}(\omega_t)]}{\gamma_{eff}}
%\end{equation}
It is possible to exploit Eq.~\ref{phonon} to estimate a maximum injected cooling power  before the fiber phase noise starts contributing significantly.

Phase noise introduced by the fiber can have a significant impact on detection sensitivity since it could increase the detection noise floor. This can be evaluated by looking at the homodyne PSD of the resonant meter field. By using Eqs.~\ref{dyn2} and by defining $K_i(\omega)=[\kappa-i(\omega+\Delta_i)]^{-1}$ and $G(\omega)=i\hbar g_o^2 k^2 K_m(\omega)\chi_{eff}(\omega)$ we can express the intra-cavity meter field as
%\begin{eqnarray}\label{intra_meter}
%\nonumber
%a_m &=& v_m~ K_m(\omega)\left[   1+  |\alpha_m|^2 G(\omega)  \right]  + v_m^\dag \left[\alpha_m^2 K_m^*(-\omega) G(\omega)    \right] \\ &+& v_c \left[ \alpha_m \alpha_c^* K_c(\omega) G(\omega)   \right]+ v_c^\dag \left[ \alpha_m \alpha_c K_c^*(-\omega)  G(\omega)  \right] \\  &+&
% \dot{\phi}_m~\alpha_m \left\{-i K_m(\omega)-G(\omega)\left[ i |\alpha_m|^2\left(  K_m^*(-\omega)-K_m(\omega) \right)\right]\right\}\\&+ & \dot{\phi}_c \left\{ -G(\omega) \alpha_m \left[ i |\alpha_c|^2  \left(  K_c^*(-\omega)-K_c(\omega) \right)   \right]  \right\} \\&+ &  i g_o k \alpha_m K_b(\omega) \chi_{eff}(\omega)~\xi
%\end{eqnarray}
\begin{equation}\label{intra_meter}
%\nonumber
\begin{split}
a_m& = v_m~ K_m(\omega)\left[   1+  |\alpha_m|^2 G(\omega)  \right]  + v_m^\dag \left[\alpha_m^2 K_m^*(-\omega) G(\omega)    \right] \\ +& v_c \left[ \alpha_m \alpha_c^* K_c(\omega) G(\omega)   \right]+ v_c^\dag \left[ \alpha_m \alpha_c K_c^*(-\omega)  G(\omega)  \right] \\  +&
 \dot{\phi}_m~\alpha_m \left\{i K_m(\omega)-i G(\omega) |\alpha_m|^2 \left[K_m(\omega)-  K_m^*(-\omega) \right]\right\}\\+ & \dot{\phi}_c \left\{ -i G(\omega) \alpha_m  |\alpha_c|^2  \left[ K_c(\omega)- K_c^*(-\omega)  \right]  \right\} \\+ &  i g_o k \alpha_m K_b(\omega) \chi_{eff}(\omega)~\xi
\end{split}
\end{equation}
\noindent By using standard input-output formalism the reflected meter field is given by $b_{out}=-a_{in,m}+\sqrt{2 \kappa_{in}}~a_m$;  than as usual the homodyne observable is defined as $\mu=b_{out}~e^{-i\theta}+b_{out}^\dag~e^{i\theta}$.

\textbf{Results}. We consider a FFP whose input coupler is held at $L_{free}=4$~mm from the fiber input face and a $100$~m long fiber at the end of which an ideal mirror is assumed. The fiber has a core (cladding) diameter of $8.7~(125)~\mu$m and a mfr$=5.25~\mu$m. The system is considered to be held in a UHV environment at a pressure $P=10^{-9}$~mbar which corresponds to a gas-damping coefficient $\gamma_g=32 P/\pi \bar{v} \rho t$. The cavity finesse is $\mathcal{F}=10$, which gives a $FSR=1$~MHz and a cavity half-linewidth $\kappa/2\pi=51$~kHz, optical losses introduced by the fiber-free space interface contribute to the overall decay channel by $\sim7\%$. The apodized microdisk  has a radius of $8~\mu$m and a thickness $t=0.5~\mu$m. With these values the coupling parameter is $g_o/2\pi=3$~MHz. The trapping frequency is chosen to be $\omega_t/2\pi=10^5$~Hz which gives a trapping beam power of $P_t=60$~mW. An estimate of the optimal cooling beam power can be obtained using Eq.~\ref{phonon} by requiring that the phase noise contribution equals the cooling beam back-action. That is, we impose $2\Delta_i^2\frac{m \omega_t}{\hbar g_o^2 k^2}~A_{c,-}S_{\dot{\phi}\dot{\phi}}(\omega_t)=1$. Assuming a detuning of $\Delta_c=-\omega_t$ and a ratio $r=\omega_t/\kappa$ we find $P_c^{max}\simeq\hbar \omega_l \frac{1+r^2}{4 r^4} \frac{\omega_t^2}{S_{\dot{\phi}\dot{\phi}}(\omega_t)}=12~\mu$W for our parameters.  With these parameters the optical cooling rate $\gamma_{opt}/2\pi=300$~Hz ($Q_{eff}\simeq330$). We consider a meter beam power of $P_m=4.3~\mu$W which provides a good compromise between final phonon number occupation and peak-to-noise ratio (PNR) in the homodyne detection.
\begin{figure}[h]
\begin{center}
  \includegraphics[width=0.47\textwidth]{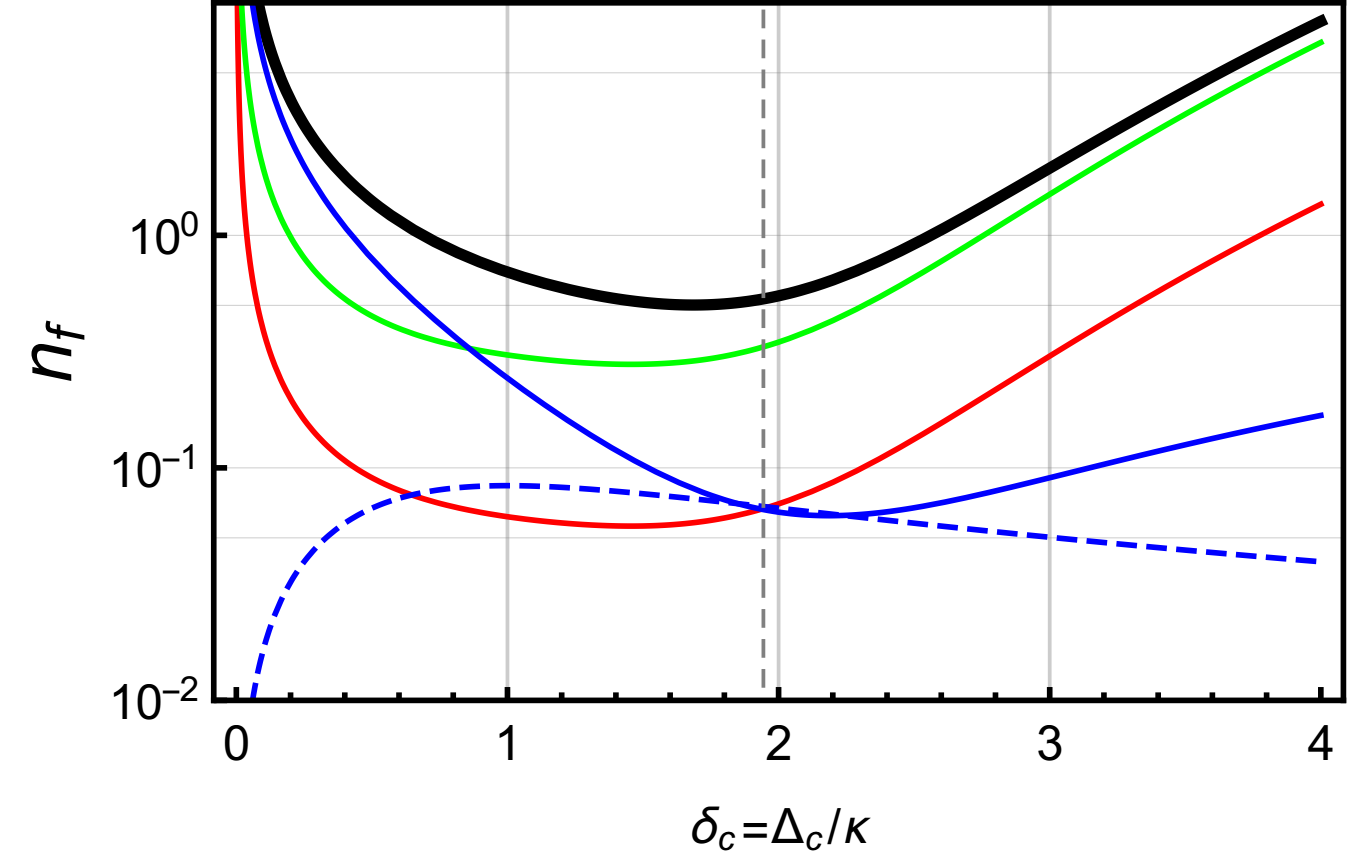}
  \caption{\label{nf} Expected final phonon number as function of the cooling field normalized detuning. All contributions are shown: total (black), thermal noise (red), cooling field back-action (blue), meter field back-action (green) and fiber phase noise contribution (dashed-blue).  The vertical dashed-gray line indicates the detuning $\Delta_c=\omega_t$ that maximizes the cooling rate in the resolved sideband regime. }
\end{center}
\end{figure}
Despite the extremely low finesse a final thermal occupation number smaller than one can be obtained. This is shown in Fig.\ref{nf} where we plot the final effective phonon number $n_f$ as a function of cooling field detuning $\delta_c$.  As imposed, fiber phase noise gives an equal contribution to the cooling field back-action, however, the limiting contribution comes from the back-action of the meter. A direct consequence is that the minimal $n_f$ is no longer obtained for the typical optimal detuning in the resolved sideband regime but at a slightly lower value. This is found to be $\delta_c=-0.87~\omega_t$ for which  a $n_f=0.5$ is obtained. Interestingly, without the meter back-action the final phonon number would be $n_f=0.17$ despite the contribution from the fiber phase noise.
\begin{figure}[h]
\begin{center}
  \includegraphics[width=0.47\textwidth]{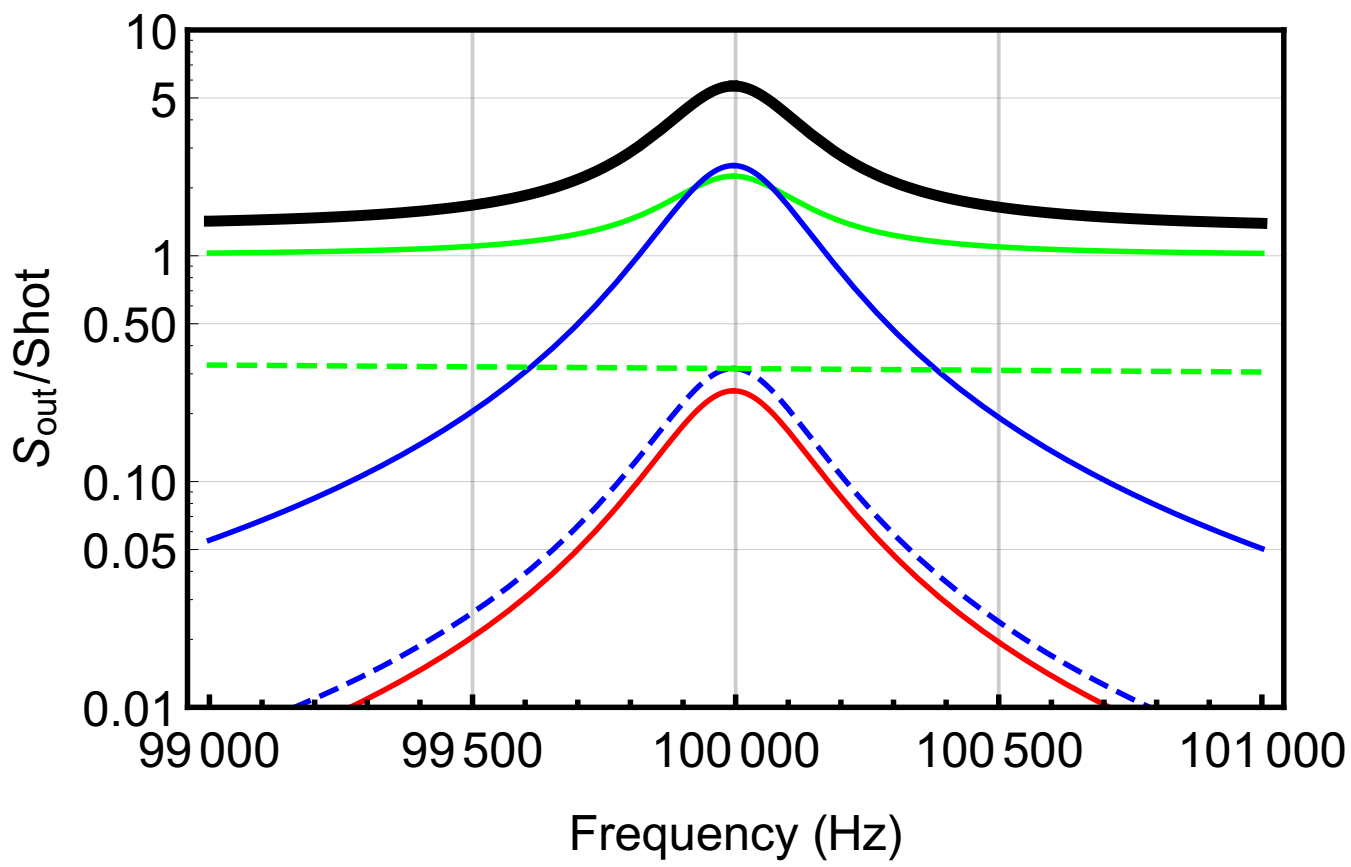}
  \caption{\label{hdyne} Phase quadrature homodyne spectra of the meter field normalize to shot noise. All contributions are shown: total (black), thermal noise (red), cooling field back-action (blue), meter field back-action (green), fiber phase noise contribution for the meter field (dashed-green) and  fiber phase noise contribution for the cooling field (dashed-blue).   }
\end{center}
\end{figure}

To verify the detectability of the microdisk motion we evaluated the homodyne spectra of the phase quadrature for the resonant meter field. This is show in Fig.~\ref{hdyne} where we plot the total quadrature PSD normalized to shot noise along with all contributions. The dominant noise floor is given by the meter field shot noise with a non-negligible contribution due to fiber phase noise. We point out that this is the case since the trapping frequency for the microdisk is significantly higher than the frequency cut-off described by Eq.~\ref{fomega}, indeed, phase noise contribution is orders of magnitude higher at low frequency.
\begin{figure}[h]
\begin{center}
  \includegraphics[width=0.47\textwidth]{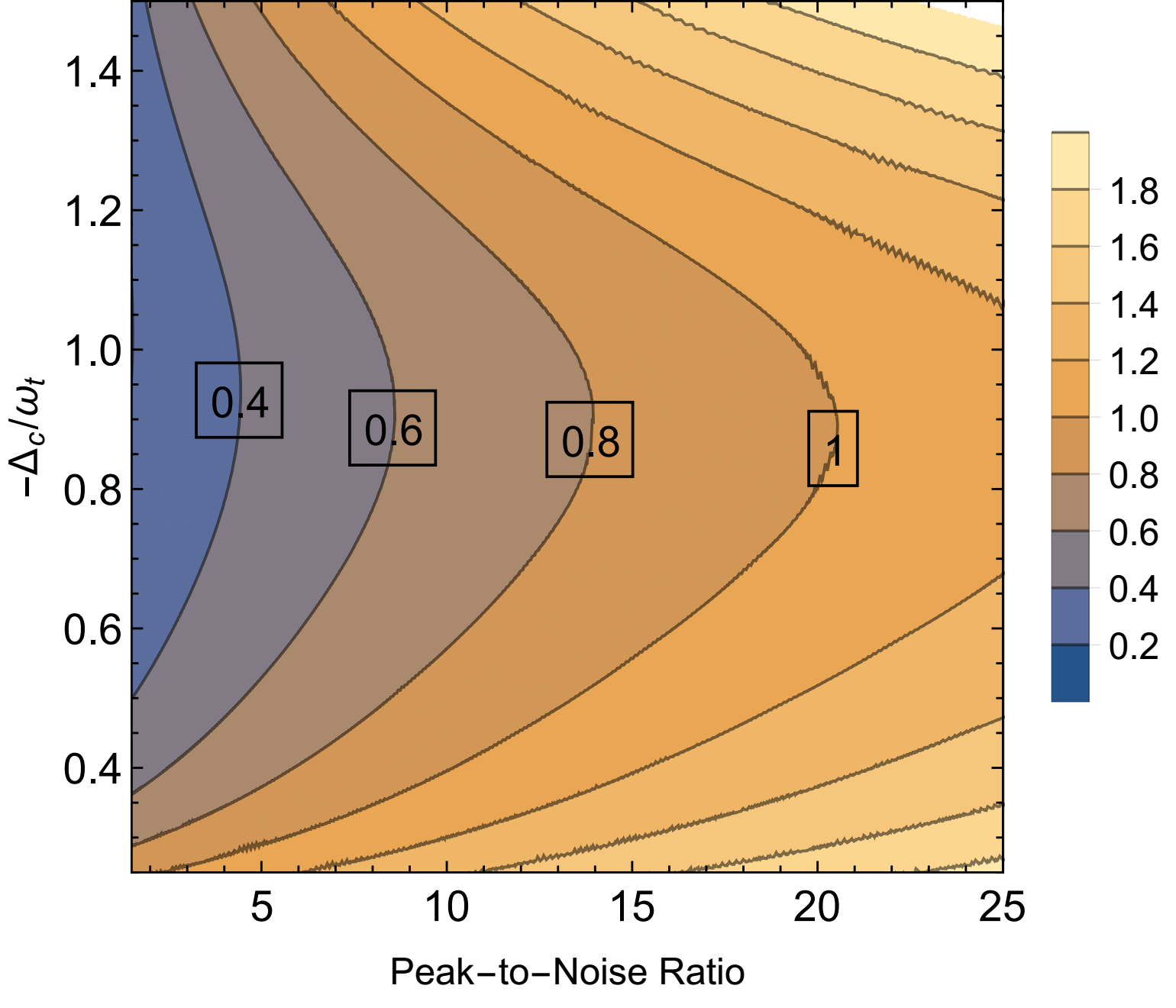}
  \caption{\label{contour} Contour plot of the achievable final thermal occupation number as a funtion of cooling beam detuning and achievable PNR.   }
\end{center}
\end{figure}

In order to emphasize the tradeoff between detectability and final occupation number, we show in Fig.~\ref{contour} a contour plot of $n_f$ as function of cooling beam detuning $\Delta_c$ and achievable peak-to-noise ratio. A final $n_f=1$  can be obtained with a high PNR$=25$ with an input power of $P_m=12.3~\mu$W. Interestingly, $n_f$ has a smooth dependance on $\Delta_c$ since the system is not deeply into the resolved sideband regime.

In conclusion, we have shown that an apodized microdisk trapped in an extrinsic Fiber Fabry-Perot interferometer could be cooled down to the quantum ground state despite the extremely low finesse of the system. Thermoptic phase noise introduced by random temperature fluctuations along the fiber has been taken into account and has been shown not to constitute an intrinsic limit toward ground state cooling. Further analysis is however required. The intra-cavity power of the trapping beam is $\sim~360$~mW, this value coincides with the threshold for Brillouin scattering for a single pass in the $100$~m long fiber considered here. This implies that additional measures to significantly increase the Brillouin threshold need to be put in place. An intriguing possibility is the use of photonic crystal hollow-core fibers (HCF) due to an increased power handling capability thanks to a reduced interaction with silica\cite{Ouzounov}. At the same time, a lower thermal phase noise level have been measured for HCFs\cite{cranch} allowing more flexibility in the parameters choice. Optical losses at the interface have already been estimated and found of the same order as for a standard single mode fiber, however, HCFs have significantly higher losses and coupling to higher modes could impact the system performance.

%AAG slightly reworded and expanded
It has been recently proposed that a levitated sensor could be exploited to detect high frequency gravitational waves \cite{geraci1}. It has been shown that, under the right conditions, the attainable sensitivity could be more than an order of magnitude better than current interferometers like LIGO and VIRGO in the frequency range of $50-300$\,kHz. The configuration considered here could represent a viable alternative to implement such a proposal, and will be studied in future work, with the fiber-based cavity potentially eliminating the demand for large optical mirrors.  A variety of sources could produce gravitational waves at such frequencies, including signals from Black Hole superradiance \cite{minasuperrad}. For example such signals can be associated with the QCD axion, a notable dark matter candidate \cite{masha1}. Such sources can also be sought after in current advanced gravitational wave interferometer observatories \cite{masha2}, and the more compact levitated-sensor approach could significantly expand the search capabilities in the higher frequency band \cite{geraci1}.

%AAG added acknowledgements
\emph{Acknowledgements} The authors acknowledge funding from the EPSRC Grant No. EP/N031105/1. AP has received funding from the European Union’s Horizon 2020 research and innovation programme under the Marie Sklodowska-Curie Grant Agreement No. 749709. AG is supported by the U.S. National Science Foundation grant no. PHY-1506431 and the Heising-Simons Foundation.

\end{document}